\documentclass[preprint,aps,superscriptaddress,nofootinbib]{revtex4}
\usepackage{amsmath}
\usepackage{graphicx}
\usepackage{color}
\usepackage{slashed}

\begin{document}

\title{Dipole Coupling Effect of Holographic Fermion in Charged Dilatonic Gravity}
\author{Wen-Yu Wen\thanks{%
E-mail: steve.wen@gmail.com}}
\affiliation{Department of Physics, Chung Yuan Christian University, Chung Li City, Taiwan}
\affiliation{Leung Center for Cosmology and Particle Astrophysics\\
National Taiwan University, Taipei 106, Taiwan}

\author{Shang-Yu Wu\thanks{%
E-mail: loganwu@gmail.com}}
\affiliation{Institute of physics, National Chiao Tung University, Hsinchu 300, Taiwan}

\begin{abstract}
In this note, we study the dipole coupling effect of holographic fermion in a charged dilatonic black hole proposed by Gubser and Rocha \cite{Gubser:2009qt}.  It is found that the property of Fermi liquid is rigid under perturbation of dipole coupling, and the Fermi momentum is linearly shifted.  A gap is dynamically generated as the coupling becomes large enough and the Fermi surface ceases to exist as the bulk dipole coupling further increases.
\end{abstract}

\maketitle

%\date{\today}

%\keywords{}

\section{Introduction and Summary}
Holographic duality, best known in the context of AdS/CFT correspondence \cite{Maldacena,Polyakov,Witten}, states that any gravity theory in a $(d+1)$-dimensional spacetime with asymptotical anti-de Sitter geometry has a dual description of quantum field theory living at its $d$-dimensional boundary.  One advantage of this duality is that several observables at the strongly coupled limit in the field theory side can be computed in the gravity side at the large radius of curvature limit, even though the knowledge of underlying microscopic degrees of freedom may still be missing.  In the past, several holographic gravity models have been proposed to capture behaviors of (non-)Fermi liquid on the boundary condensed matter system \cite{Liu:2009dm,Gubser:2009qt,Faulkner:2009wj,Iqbal:2011ae}.  Among them, a charged dilatonic black hole proposed by Gubser and Rocha \cite{Gubser:2009qt} is particularly attractive for two features: at first, different from other bottom-up approaches, this background can be seen as the dimensional reduction from the near-horizon geometry of three-equal-charge black hole in $AdS_4\times S^7$ \cite{Cvetic:1999xp}.  Secondly, it was claimed that in the infrared this model is dual to a Fermi liquid thanks to its linear relation between entropy and temperature.  Normal modes of a probed fermion were studied in the extremal limit of charged black hole \cite{Gubser:2009qt} and linear dispersion relation was verified in \cite{Wu:2011cy}.

On the other hand, a boundary theory dual to the AdS Reissner-Nordstr\"{o}m black hole in the presence of a dipole interaction term is proposed \cite{Edalati:2010ww} and studied in several following works \cite{Edalati:2010ge,Guarrera:2011my,Li:2011nz,Wu:2012fk}.  Upon tuning the dipole coupling strength, the boundary theory is found in a phase of either Fermi liquid, non-Fermi liquid or Mott insulator.  We remark that while the extremal AdS RN background, without dipole coupling, was first claimed to be dual to a boundary theory of non-Fermi liquid at quantum critical point \cite{Liu:2009dm}, it is not clear which role does the dipole coupling play in a boundary theory of Fermi liquid.  It is the goal of this paper to explicitly investigate the dipole coupling effect on a particular boundary theory of Fermi liquid, which is dual to the charged dilatonic black hole by Gubser and Rocha.

At last, we summarize our finding as follows:
\begin{itemize}
\item We have found the structure of Fermi surface constructed by Gubser and Rocha is {\sl rigid} regardless the presence of dipole coupling. There is no sign about the Fermi/non-Fermi liquid transition by varying bulk dipole coupling.
\item Fermi momentum can be linearly shifted by the dipole coupling strength.
\item A gap appears for large dipole coupling strength, as a generic feature of dipole interaction. However, the gap might not be the Mott one,  since we could not observe any clear sign of spectral weight transfer from the spectral function.
\item The gap persists at finite temperature for large enough dipole coupling.
\end{itemize}

In the next two sections, we briefly describe the geometry of dilatonic black hole constructed in \cite{Gubser:2009qt}, and Dirac equation in this background. In section 4, we show the rigidity of Fermi surface of this background under the bulk dipole coupling. In section 5, we show the emergence of a gap at zero and finite temperature. Conclusions and discussions are addressed in the end.

\section{Preparing a Fermi liquid}
In this section, we review Gubser and Rocha's construction of a charged dilatonic black hole, which is claimed dual to a boundary theory of Fermi liquid.  The three-equal-charge black hole has the following lagrangian \cite{Gubser:2009qt}:
\begin{equation}
{\cal L} = \frac{1}{2\kappa^2} \bigg[ R - \frac{1}{4}e^{\alpha}F^2_{\mu\nu}-\frac{3}{2}(\partial_\mu \alpha)^2 + \frac{6}{L^2}\cosh{\alpha} \bigg],
\label{eq:background action}
\end{equation}
and the charged black hole solution is given by
\begin{eqnarray}
&&ds^2 = e^{2A}(-h dt^2 + d\vec{x}^2) + \frac{e^{-2A}}{h}dr^2,\qquad F=dA,\nonumber\\
&&A = \log{\frac{r}{L}}+\frac{3}{4}\log{\big( 1+\frac{Q}{r} \big)},\qquad  h = 1-\frac{\nu L^2}{(Q+r)^3},\nonumber\\
&&\alpha = \frac{1}{2}\log{\big( 1+\frac{Q}{r} \big)}, \qquad A_t = \frac{\sqrt{3Q\nu}}{Q+r}-\frac{\sqrt{3Q}\nu^{\frac{1}{6}}}{L^{\frac{2}{3}}}
\label{background}
\end{eqnarray}
This solution is asymptotic to AdS, but the IR geometry is neither Lifshitz nor AdS.
The zero temperature is obtained by taking extremal limit where $\nu=\frac{Q^3}{L^2}$ with vanishing horizon $r_0=0$. The temperature and boundary chemical potential are given by \cite{Li:2011sh}
\begin{equation}
T=\frac{3\nu^{\frac{1}{6}}}{4\pi L^{\frac{5}{3}}}\sqrt{r_{0}}, \quad \mu=\frac{\sqrt{3Q}\nu^{\frac{1}{6}}}{L^{\frac{2}{3}}}.
\end{equation}
It was also shown that at low temperature the entropy density $\hat{s}$ is proportional to temperature,
\begin{equation}
s \approx -\frac{8\sqrt{3}\pi^{3} L}{3\kappa^{2}}\mu T,
\end{equation}
where $\kappa^{2}=8\pi G$. This implies a boundary theory of Fermi liquid dual to this gravity background.
Later it was explicitly checked that near Fermi surface $k_F$, the linear relation
\begin{equation}
\omega \propto (k-k_F)
\end{equation}
holds \cite{Wu:2011cy}.

\section{Probed fermion with dipole coupling}
Now we would like to {\it dope} the system with a dipole coupling term in the bulk.  The main purpose of this paper is to study the dipole effect to the underlying Fermi liquid system.  We consider the bulk spinor action:
\begin{equation}
{\cal S}_D = i \int{d^4x \sqrt{-g}\bar{\psi}(\slashed{\cal{D}}- m - ip \slashed{F})\psi},
\end{equation}
where the flat space Gamma matrices $\Gamma^a = e_{\mu}^{a}\Gamma^{\mu}$, $\slashed{F}=\frac{1}{2}\Gamma^{\mu\nu}(e_{\mu})^{a}(e_{\nu})^{b}F_{ab}$, $\Gamma^{\mu\nu}=\frac{1}{2}[\Gamma^{\mu},\Gamma^{\nu}]$, $(e_{\mu})^{a}$ is the vielbein and covariant derivative $\slashed{\cal D}=\Gamma^{a}D_{a}=\partial_a + \frac{1}{4}(\omega_{\mu\nu})_a\Gamma^{\mu\nu}-iqA_a$.
The bulk Dirac equation is
\begin{equation}
(\slashed{D}-m-ip\slashed{F})\psi=0.
\end{equation}
Taking the ansatz $\psi = (-gg^{rr})^{\frac{1}{4}}e^{-i\omega t +i k_{i}x^{i}}\varphi$, one can remove the spin connection in the Dirac equation. To decouple the equations of motion, we introduce projectors $\Gamma_\pm=\frac{1}{2}(1\pm \Gamma^{r}\Gamma^{t}\hat k\cdot\vec\Gamma)$ and write $\varphi_{\pm}=\Gamma_{\pm}\varphi(r)$.
With a choice of Gamma matrices
\begin{equation}
\Gamma^{r}=
\begin{pmatrix}
-\sigma_{3} & 0\\
0 & -\sigma_{3}
\end{pmatrix}, \quad
\Gamma^{t}=
\begin{pmatrix}
i\sigma_{1} & 0\\
0 & i\sigma_{1}
\end{pmatrix}, \quad
\Gamma^{1}=
\begin{pmatrix}
-\sigma_{2} & 0\\
0 & \sigma_{2}
\end{pmatrix}, \quad
\Gamma^{2}=
\begin{pmatrix}
0 & -i\sigma_{2}\\
i\sigma_{2} & 0
\end{pmatrix},
\end{equation}
one obtains the bulk Dirac equation
\begin{equation}
(e^{2A}\sqrt{h}\sigma^{3}\partial_{r}+me^{A}+ipe^{A}A'_{t}\sigma^{2})\varphi_{\pm}+(-\frac{\sigma^{1}}{\sqrt{h}}(\omega+qA_{t})\pm ik\sigma^{2})\varphi_{\pm}=0.
\label{Dirac equation}
\end{equation}
Where we use the rotational symmetry in the spatial directions to set $k_{x}=k$, and $k_{i}=0$ for $i\neq x$.
Set $\varphi_{\pm}=\begin{pmatrix} y_{\pm} \\ z_{\pm} \end{pmatrix}$, and define $\xi_{\pm}=\frac{y_{\pm}}{z_{\pm}}$, one can obtain the flow equations
\begin{equation}
e^{2A}\sqrt{h}\partial_r\xi_\pm +2me^{A}\xi_\pm + (-\frac{1}{\sqrt{h}}(\omega+qA_{t})\mp k-pe^{A}A'_{t})\xi_\pm^2+(pe^{A}A'_{t}-\frac{1}{\sqrt{h}}(\omega+qA_{t})\pm k)= 0.
\label{eom_flow}
\end{equation}
This equation is solved by imposing an in-falling condition at the horizon, such that
\begin{equation}
\xi_\alpha = i \quad \text{as} \quad  r\to r_h \quad (\omega \neq 0)
\end{equation}
\begin{figure}
\center{
\includegraphics[width=6cm]{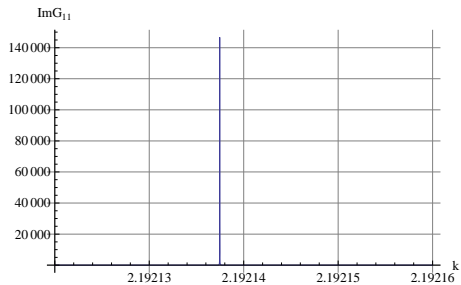}\hspace{0.2cm}
\caption{\label{fermi momentum} The plot for $ImG_{11}(k)$ at $\omega=-10^{-10}$ in the absence of dipole coupling $p$. There is a sharp quasiparticle peak that indicates a Fermi surface. The fermi momentum is found at $k=2.19213747$.}
}
\end{figure}
We remark that the in-falling condition is not modified by the dipole coupling thanks to its subleading contribution near the horizon.  The boundary retarded Green function can be extracted by
\begin{equation}
G_{R}=
\begin{pmatrix}
G_{+} & 0\\
0 & G_{-}
\end{pmatrix}
\end{equation}
where
\begin{equation}
G_\pm = \lim_{r\to \infty} r^{2m}\xi_\pm.
\label{eom_green}
\end{equation}
Note the Dirac equation eq.(\ref{Dirac equation}) implies the relation $G_{11}(\omega,k)=G_{22}(\omega,-k)$. So in the following sections, we will only show the 3D plots and density plots for $Im G_{22}$ and also set $q=2$, $L=1$ ,$Q=1$ and $m=0$ for convenience.
\section{Rigidity of Fermi surface}
It is shown that a simple holographic realization of finite density fermionic system exhibits single particle spectral function with sharp Fermi surface of a form distinct from the Landau-Fermi liquid \cite{Liu:2009dm,Faulkner:2009wj}. Furthermore, it is observed that in this finite density background (RN-AdS) \cite{Edalati:2010ge}, one can find Fermi or non-Fermi liquid behavior by tuning the bulk dipole coupling strength $p$, and then the sharp Fermi surface ceases to exist for large $p$. Motivated by this, it is interesting to investigate if this feature is robust or not in general backgrounds.
So we would like to consider the bulk dipole coupling effect in the charged dilatonic black hole in \cite{Gubser:2009qt}. Similar to RN-AdS case, we actually find a sharp quasiparticle peak around $\omega=0$ which indicates a Fermi surface (see figure (\ref{fermi momentum})), and the dipole coupling does not change the structure of Fermi surface but just shift the location of Fermi momentum.  However, the height of peak dramatically decreases for large $p$ and is too small to be identified as a Fermi surface.  For instance, the peak of $ImG_{11}$ drops under $100$ for $p=4$ but of order of unity for $p=10$.

\begin{figure}
\center{
\includegraphics[width=5cm]{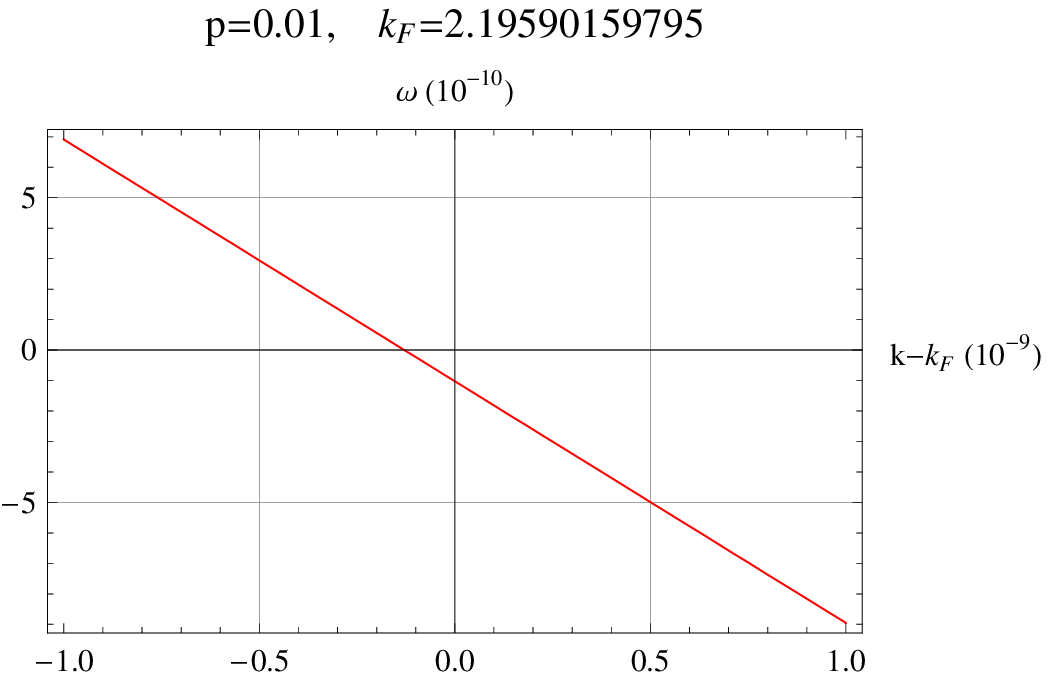}\hspace{0.2cm}
\includegraphics[width=5cm]{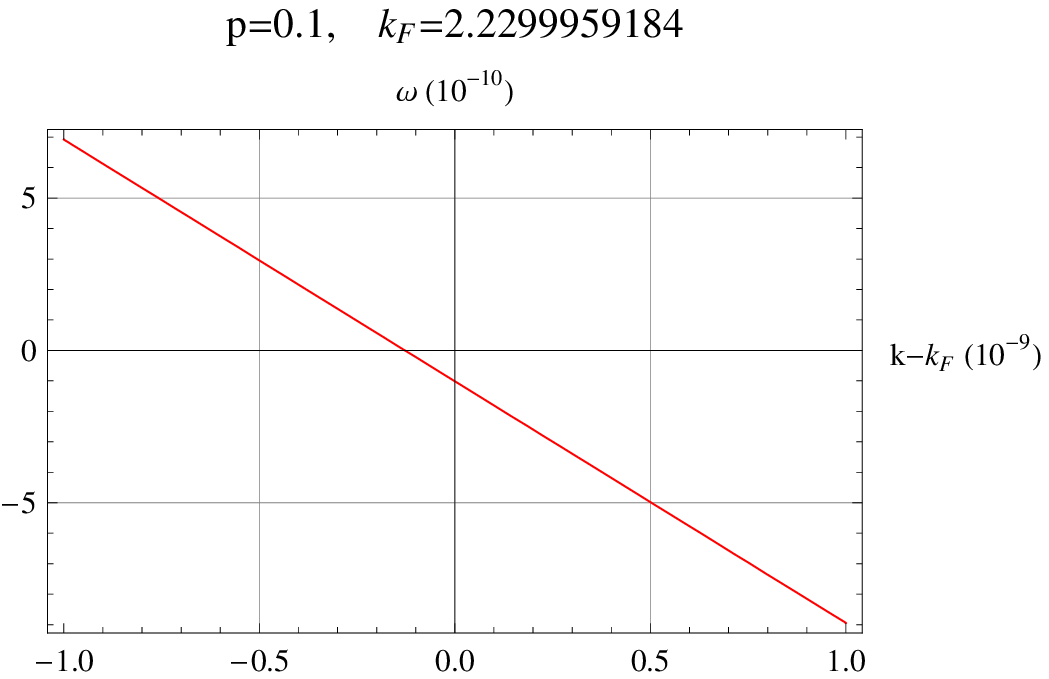}\hspace{0.2cm}
\includegraphics[width=5cm]{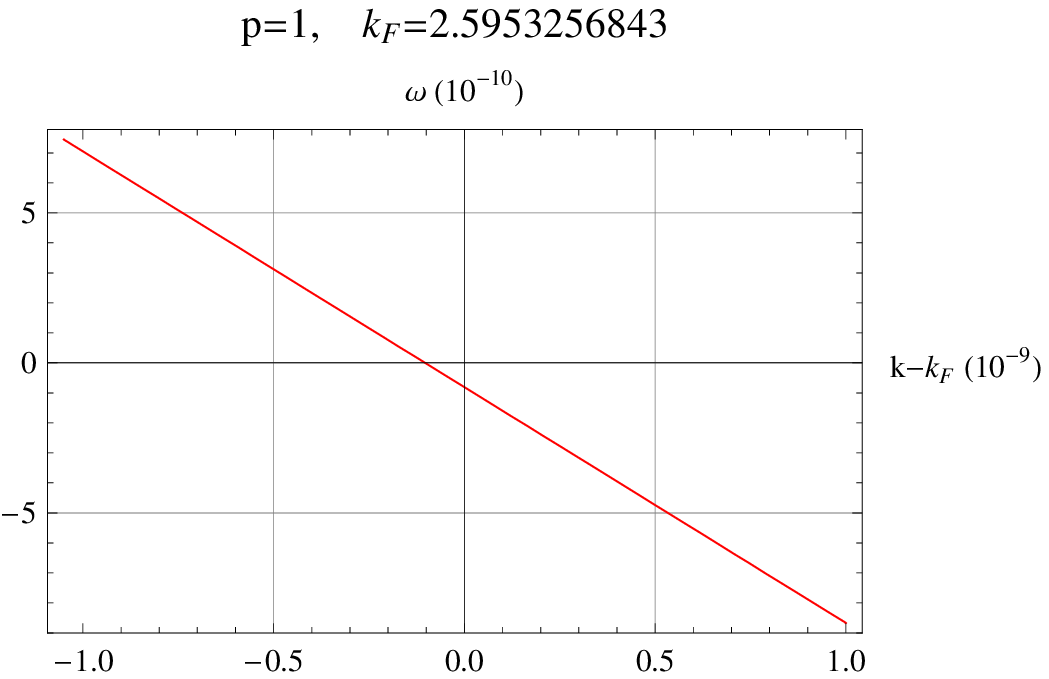}\hspace{0.2cm}
\includegraphics[width=5cm]{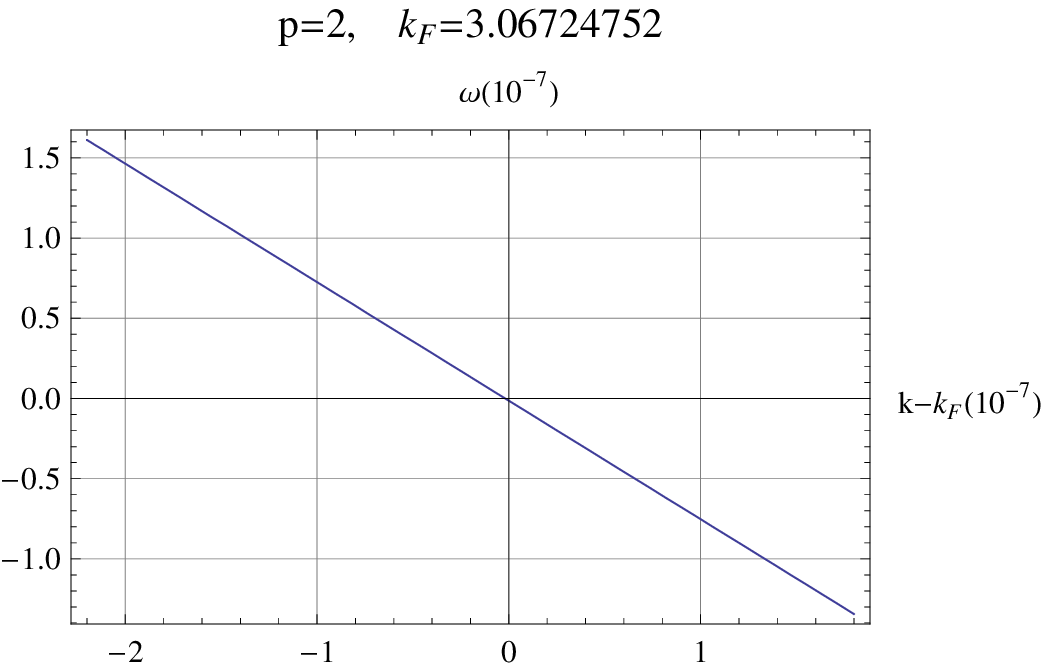}\hspace{0.2cm}
\includegraphics[width=5cm]{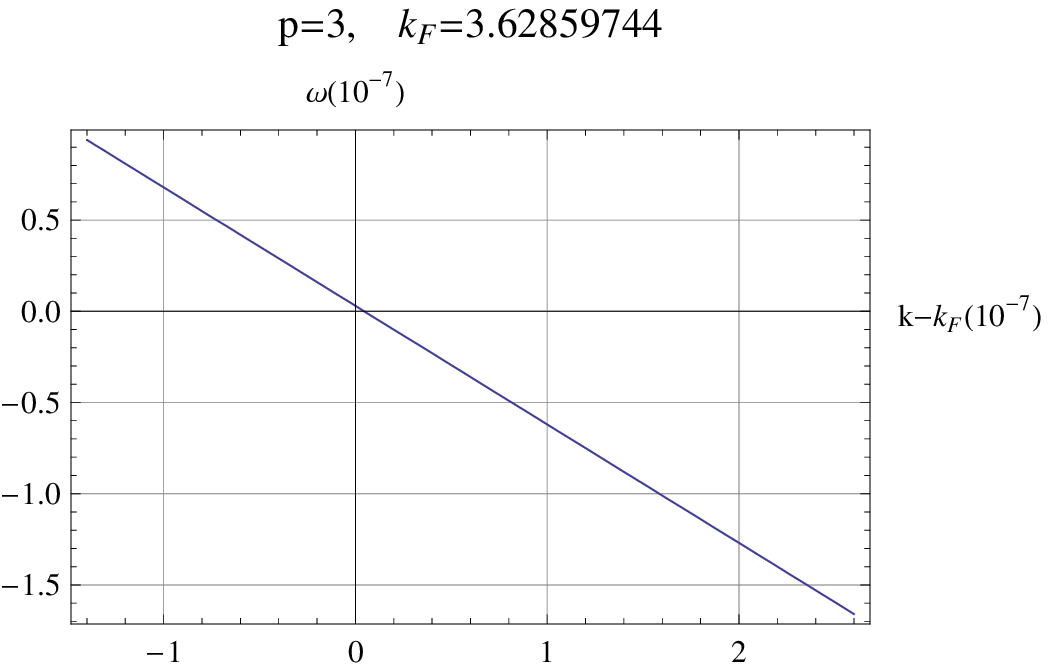}\hspace{0.2cm}
\includegraphics[width=5cm]{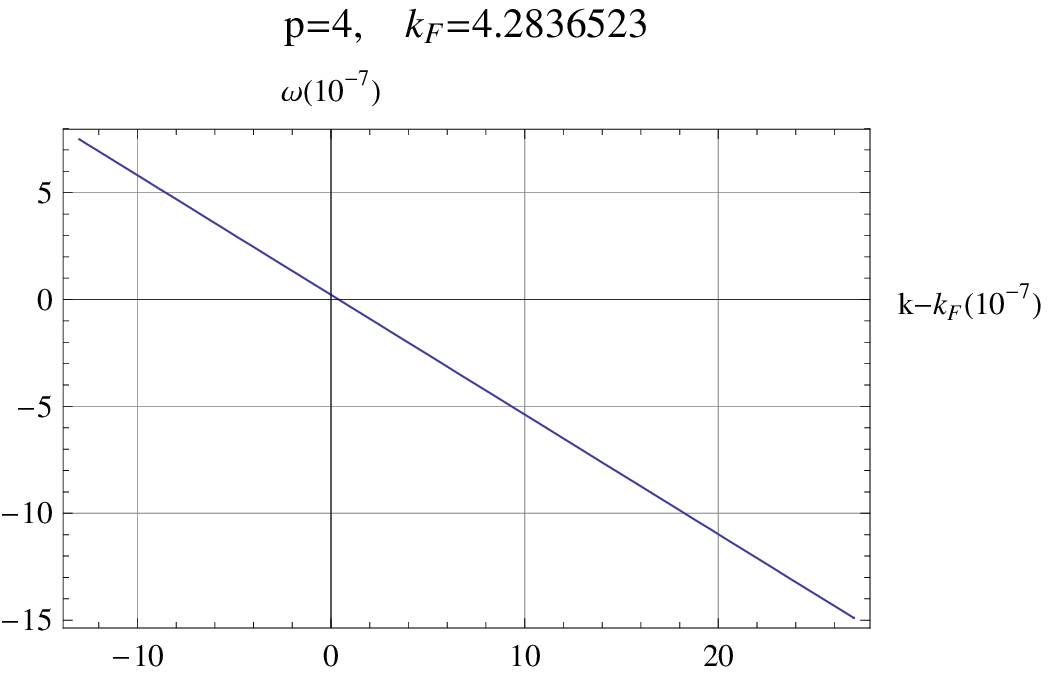}\hspace{0.2cm}
\caption{\label{fig:dispersion} Dispersion relation near the Fermi surface for $p=0.001,0.1,1,2,3,4$.  We remark that the Fermi momentum $k_F$ linearly increases with dipole coupling strength $p$, and linear dispersion relation holds for generic $p$.}
}
\end{figure}

We do not observe any sign of non-Fermi liquid numerically by tuning the dipole coupling strength $p$. In the figure (\ref{fig:dispersion}), we show the dispersion relation near the Fermi surface for various $p$. The linear relation between $k-k_F$ and $\omega$ has a universal form:
\begin{equation}
\omega \simeq v_g \text{ } (k-k_F) + \text{const},
\end{equation}
which implies the Fermi surface constructed in \cite{Gubser:2009qt} is rigid and the property of Fermi liquid remains regardless of the bulk dipole coupling. The group velocity $v_g=\partial\omega / \partial k$ can also be read off as $v_g\simeq 0.79 c$ for $p \le 1$ and slows down for larger $p$, say $v_g\simeq 0.5c$ for $p=5$.  

\begin{figure}
\center{
\includegraphics[width=6cm]{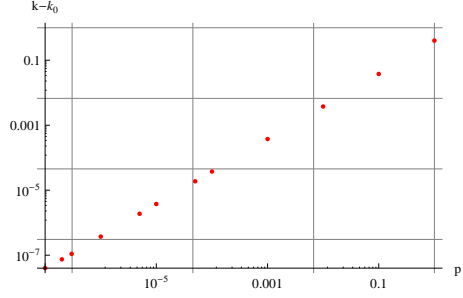}\hspace{0.2cm}
\caption{\label{fig:pk} The shift of Fermi momentum is plotted against various dipole coupling strength $p$.  The best fit shows the shift of Fermi momentum $k-k_0 \propto 0.66103 \text{ } p^{1.00161}$, where $k_0=2.19213747$ is the pole of retarded Green function $G_{11}$ at $\omega = - 10^{-10}$ and identified as the Fermi momentum at $p=0$.}
}
\end{figure}

According to the numerical result, for massless Dirac fermion and zero frequency, we can see the pole of Green function is shifted linearly up to some value of $p$, before the whole system enters the insulator phase where the pole disappears. In the figure (\ref{fig:pk}), the shift of Fermi momentum is plotted against various dipole coupling strength $p$. The numerical result shows the best fit for relation:
\begin{equation}
\label{linear_relation}
k-k_0 \propto 0.66103 \text{ } p^{1.00161},
\end{equation}
where $k_0=2.19213747$ is the numerically-found Fermi momentum at $p=0$. We also find there exists some negative $p$ at which the Fermi momentum goes through zero, signifying that the excitations change over from "particle-like" to "hole-like", as similar to RN-AdS case.
In the following, we show that analytical arguments, by examining the retarded Green function for massless Dirac field at both near horizon and asymptotic limit, may shed some light to understand above-mentioned numerical results: 
\subsection{Near horizon analysis}
Taking the near horizon limit, where $r\to 0$ for extremal case, one obtains the following asymptotic equation from (\ref{eom_flow}) and (\ref{eom_green}):
\begin{equation}
\frac{\sqrt{3}Q}{L^2}\partial_r G_\pm = \bigg[ \sqrt{\frac{Q}{3}}\omega r^{-3/2}\pm \frac{k}{r} + p\frac{\sqrt{3}Q^{3/4}}{L^2}r^{-3/4} \bigg]G_\pm{}^2 + 
\bigg[ \sqrt{\frac{Q}{3}}\omega r^{-3/2}\mp \frac{k}{r} - p\frac{\sqrt{3}Q^{3/4}}{L^2}r^{-3/4} \bigg].
\end{equation}
Here we intended to keep all subleading terms in order to see that the term with dipole coupling $p$ is least dominated at $r\to 0$ limit.  This implies that $p$ plays insignificant role in the IR physics, therefore the property of Fermi liquid is hardly affected by the dipole coupling. 

\subsection{Asymptotic analysis}
Taking the asymptotic limit, where $r\to \infty$, one obtains the following equation instead:
\begin{equation}
\partial_r G_{\pm} = r^{-2m-2} \bigg[ \omega -\sqrt{3}qQ - (\frac{\sqrt{3}pQ^2}{L^2} \mp k)  \bigg] G_{\pm}{}^2 + r ^{2m-2}\bigg[ \omega -\sqrt{3}qQ + (\frac{\sqrt{3}pQ^2}{L^2} \mp k) \bigg].
\end{equation}
One can immediately see that this differential equation is invariant under the shift symmetry:
\begin{equation}
k \to k + \epsilon, \qquad p \to p \pm \frac{\epsilon}{\sqrt{3}Q^2}.
\end{equation}
Though this shift symmetry is not exact in the sense that it is only respected in the asymptotic region, it still well manifests as the linear relation found numerically in equation (\ref{linear_relation}).

At last, it is not difficult to show that at both limits, the retarded Green function involves the Tangent function.  In particular, for $r\to \infty$, we have 
\begin{eqnarray}
&&G_+ = \sqrt{\frac{\nu_+}{\nu_-}}\tan{(c\sqrt{\nu_+\nu_-}-\frac{\sqrt{\nu_+\nu_-}}{r})},\nonumber\\
&&\nu_\pm = \omega -\sqrt{3}qQ \pm (\frac{\sqrt{3}pQ^2}{L^2}\mp k),
\end{eqnarray}
which is a semi-periodic function over full range of $\omega$, with a discontinuous jump at some $\omega$ (the root of $\nu_-=0$) and constant of integration $c$.  One is tempted to compare this periodicity of retarded Green function with the log-periodicity appeared in the RN-AdS case\cite{Liu:2009dm}.

\begin{figure}
\center{
\includegraphics[width=6cm]{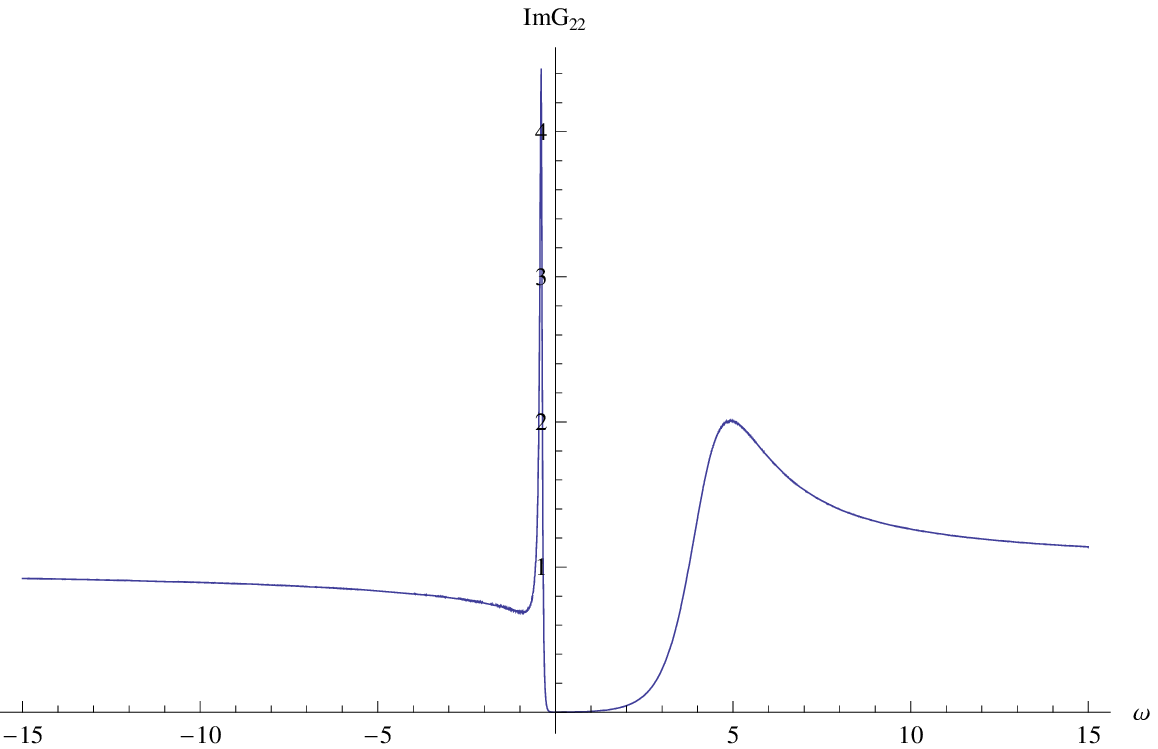}\hspace{0.2cm}
\includegraphics[width=6cm]{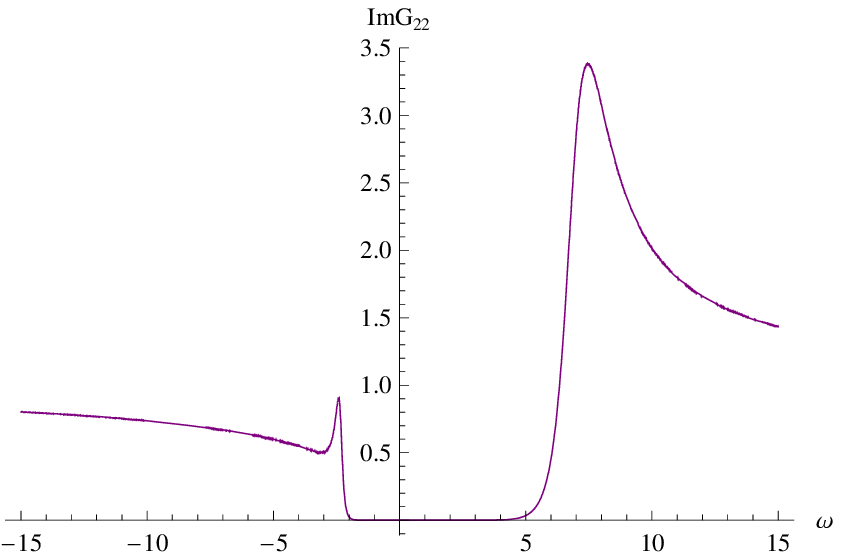}\\ \hspace{0.2cm}
\caption{\label{p0} Left Plot: The spectral function $ImG_{22}$ at $k=1.5$ ($<$Fermi momentum$\approx2.19213747$) for $p=0$. ; Right plot: The spectral function $ImG_{22}$ at $k=4$ ($>$Fermi momentum) for $p=0$.}}
\end{figure}

\section{Emergence of gap}
As it was first observed in \cite{Edalati:2010ww}, the bulk dipole coupling plays some role of Mott physics. A gap is formed for large coupling strength in the insulator phase. In addition to the formation of a gap, the spectral weight also transfers from one band to another band. We will show an emergent gap in the background studied in \cite{Gubser:2009qt}.
\subsection{Zero temperature}
Before turning on the dipole coupling, there is a sharp peak in the negative frequency, and one quasiparticle peak in the positive frequency which is more dispersive than the negative frequency one\footnote{Follow the convention in \cite{Edalati:2010ww}, we also call the band in the positive frequency as upper band and the band in the negative frequency as lower band.} (see Fig. \ref{p0}). This is contradiction with RN-AdS case in which there is only one quasiparticle peak. Moreover, the height of peak in the lower band is much higher than the one in upper band when the momentum smaller than the Fermi momentum in the absence of dipole coupling.

\begin{figure}
\center{
\includegraphics[width=6cm]{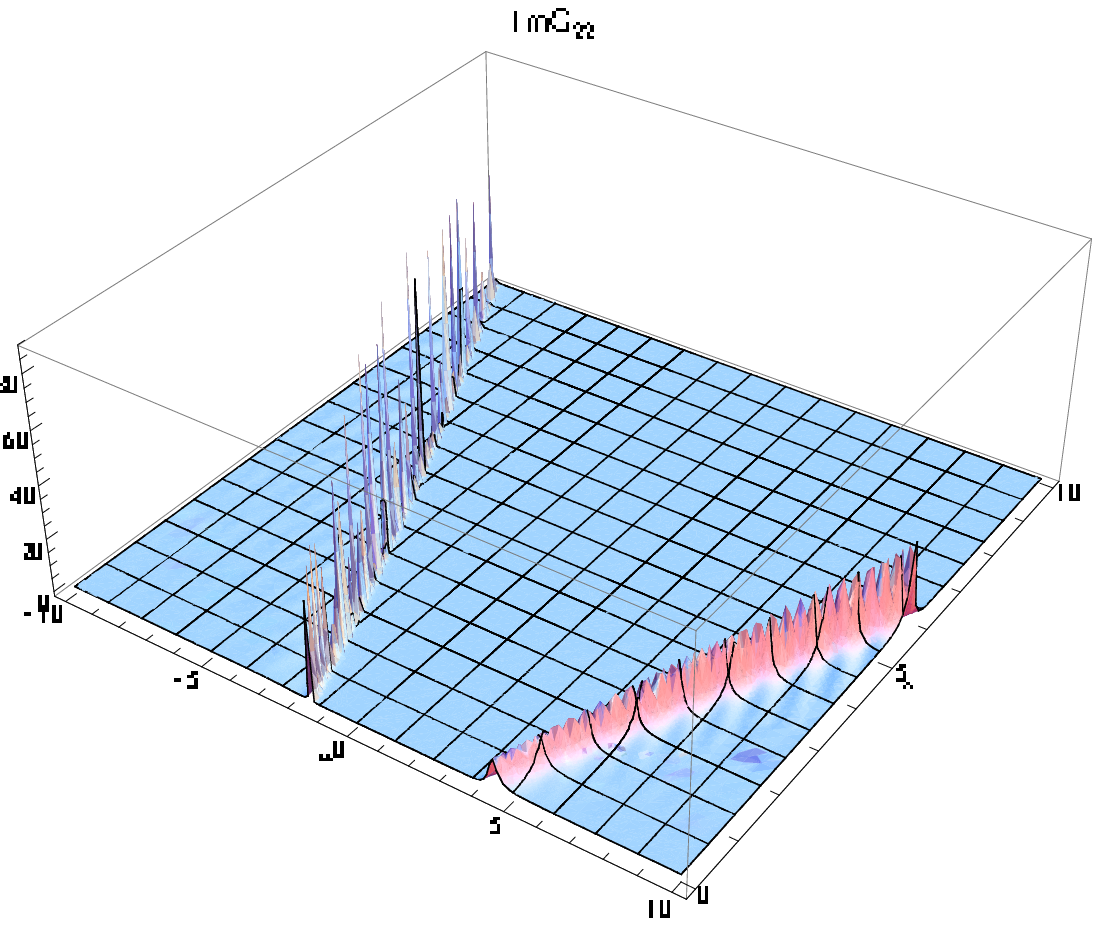}\hspace{0.2cm}
\includegraphics[width=6cm]{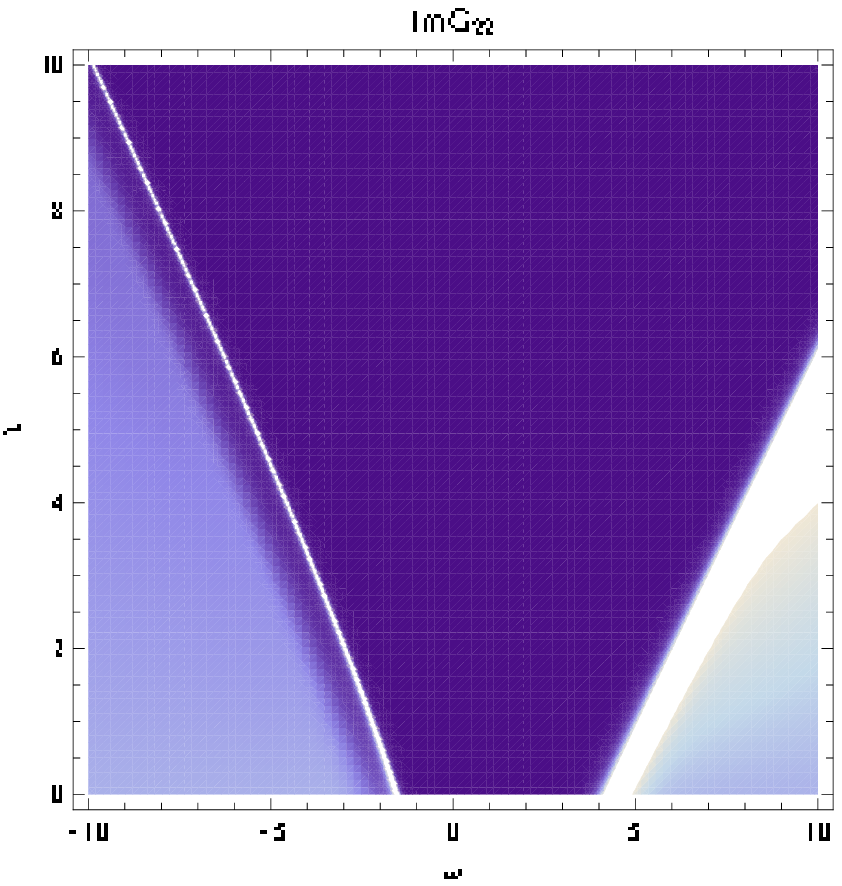}\\ \hspace{0.2cm}
\caption{\label{p4} Left Plot: The 3D plot of $ImG_{22}$ for $p=4$, Right plot: The density plot of $ImG_{22}$ for $p=4$. A gap is clearly visible for large dipole coupling.}}
\end{figure}

Turning on the dipole coupling, we find that when $p$ is larger than some critical value $p_{c}$ ($p_{c}\approx4$ in this case), a gap between two peaks opens (see Fig. \ref{p4}). Besides, as we increase $p$, the width of gap is larger and the height of both peaks becomes higher. This is different than the case in RN-AdS, where the original peak degrades and the spectral density starts to appear in the other band, namely the phenomena of spectral weight transfer happens. As $p$ is higher, the peak ceases to exist indicating the Fermi surface has disappeared. So the resulting phase is insulator phase.

\subsection{Finite temperature}
\begin{figure}
\center{
\includegraphics[width=7cm]{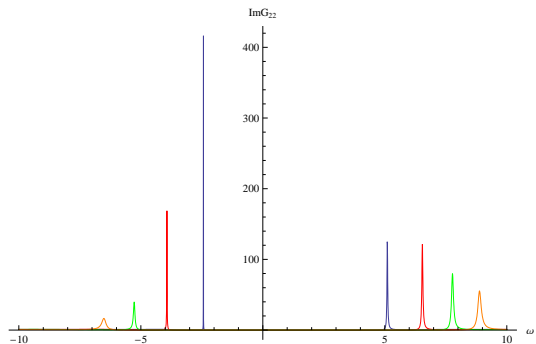}\hspace{0.2cm}}
\caption{\label{finite T}$Im G_{22}$ for $p=6$ at finite temperature. Blue: $T=0$, Red: $T/\mu=0.0974621$, Green: $T/\mu=0.137832$, Orange: $T/\mu=0.168809$. The critical value of ratio $T/\mu$ is higher than the one in RN-AdS. Besides, the width of gap increases as $T/\mu$ increases.}
\end{figure}

As shown in \cite{Edalati:2010ge}, the sharp Fermi surface is smoothed at finite temperature. Also the Mott gap generated by the bulk dipole coupling will vanish above some critical temperature. The critical value of the ratio $T/\mu$ is small ($\approx 3.92\times 10^{-2}$) for RN-AdS black hole. However, we find that the critical value of ratio $T/\mu$ in the dilatonic black hole studied in \cite{Gubser:2009qt} is higher than the one in RN-AdS. From figure \ref{finite T}, the dynamically generated gap still exists at the finite temperature for large enough dipole coupling. The width of gap also increases as the dipole coupling increases. On the other hand, the critical value of $T/\mu$ is higher for larger $p$.
\section{Discussion}
In the holographic model of non-Fermi liquid \cite{Guarrera:2011my}, it was found that the bulk dipole coupling changed the
attainable low-energy scaling dimensions, as well as the locations of the Fermi momentum. The structure of non-Fermi liquids is, however, robust under this deformation.  On the other hand, it was also found that one could find (non-)Fermi liquid behavior in different range of coupling strength by varying the coupling strength and a Mott gap is dynamically generated by the bulk dipole coupling for large dipole coupling \cite{Edalati:2010ge,Edalati:2010ww}. Moreover, the dynamically generated gap will close when the temperature is turned on and above some critical value. Conclusively, a phase diagram similar to the one of high-$T_{c}$ cuprate superconductor was found.

In this note, we have investigated a particular holographic model of Fermi liquid, proposed by Gubser and Rocha \cite{Gubser:2009qt}. Similar to the non-Fermi liquid model, we find the bulk dipole coupling change the location of Fermi momentum and the Fermi surface structure is robust under this deformation. Interestingly, it was also found that the dipole coupling linearly shifted the locations of the Fermi momentum, and we did not find any signature of non-Fermi liquid. With increasing coupling strength, the mobility of quasiparticle is decreasing and a dynamically generated gap is  observed. As the coupling strength is getting higher, the Fermi surface ceases to exist, and the system ends in an insulator phase. 

In the holographic models of non-Fermi liquid, the near horizon of extremal RN-AdS black hole is $AdS_{2}$, indicating the IR physics is well captured by an emergent $1+1$ dimensional CFT. Besides, it is argued that the IR scaling dimension of probe fermion plays an important role in the Fermi surface structure. One can apply some semi-analytical method to obtain the analytical properties of retarded Green's function in low energy limit. However, there seems a big difference from those models that the class of quantum critical point (QCP) we consider here connects Fermi liquid phase to insulator phase and does not have an explicit CFT description\footnote{We remark that while this background described by Gubser and Rocha\cite{Gubser:2009qt} does not have a $AdS$ near-horizon geometry, its lifted eleven dimensional three-equal-charge solution at extremal limit does have a $AdS_3$ near-horizon factor, which is claimed to be dual to a non-chiral CFT and responsible for the linear specific heat.}. So the semi-analytical arguments used in the non-Fermi liquid case might not apply to this case. It would be interesting to have an analytical understanding of the low energy behavior of excitations around Fermi surface by performing the matching calculations.

It would be also interesting to investigate other holographic models of QCP which associate with (non-)Fermi liquid and possibly classify various types of QCP. For example, it is shown that in the charged dilatonic black hole with Liouville potential, one can find both Fermi and non-Fermi liquid behavior when we tune the nonminimal coupling strength and dilaton potential \cite{Iizuka:2011hg}.  Another type of QCP is associated to Lifshitz-like theory\footnote{A few works studying probe fermion in Lifshitz spacetime appear recently \cite{Gursoy:2011gz,Alishahiha:2012nm,Fang:2012pw}.}.  We hope to explore the effect of dipole coupling in those backgrounds and report the results in the future\cite{future work}. 

\begin{acknowledgments}
This work is supported in part by the Taiwan's National Science Council and the National Center for
Theoretical Science.
\end{acknowledgments}

%%%%%%%%%%%%%%%%%%%%%%%%%%%%%%%%%%%%%%%%%%%%%%%%%%%%%%%%%%%%%%%


\begin{thebibliography}{99}

%%%% Gubser & Rocha's gravity
%\cite{Gubser:2009qt}
\bibitem{Gubser:2009qt}
S.~S.~Gubser and F.~D.~Rocha,
  ``Peculiar properties of a charged dilatonic black hole in AdS$_5$,''
  Phys.\ Rev.\  D {\bf 81}, 046001 (2010)
  [arXiv:0911.2898 [hep-th]].
  %%CITATION = PHRVA,D81,046001;%%

%%%% AdS/CFT correspondence
\bibitem{Maldacena} J. Maldacena,
``The Large N Limit of Superconformal Field Theories and Supergravity,''
Adv. Theor. Math. Phys. \textbf{2}, 231 (1998).

\bibitem{Polyakov} S. S. Gubser, I. R. Klebanov and A. M. Polyakov,
``Gauge Theory Correlators from Noncritical String Theory,''
Phys. Lett. \textbf{B 428}, 105 (1998).

\bibitem{Witten} E. Witten,
``Anti De Sitter Space And Holography,''
Adv. Theor. Math. Phys. \textbf{2}, 253 (1998).

%%%% Non-Fermi liquid
%\cite{Liu:2009dm}
\bibitem{Liu:2009dm}
  H.~Liu, J.~McGreevy and D.~Vegh,
  ``Non-Fermi liquids from holography,''
  Phys.\ Rev.\  D {\bf 83}, 065029 (2011)
  [arXiv:0903.2477 [hep-th]].
  %%CITATION = PHRVA,D83,065029;%%

%\cite{Faulkner:2009wj}
\bibitem{Faulkner:2009wj}
  T.~Faulkner, H.~Liu, J.~McGreevy and D.~Vegh,
  ``Emergent quantum criticality, Fermi surfaces, and AdS(2),''  Phys.\ Rev.\ D {\bf 83}, 125002 (2011)  [arXiv:0907.2694 [hep-th]].  %%CITATION = ARXIV:0907.2694;%%

%\cite{Iqbal:2011ae}
\bibitem{Iqbal:2011ae}
  N.~Iqbal, H.~Liu and M.~Mezei,
  ``Lectures on holographic non-Fermi liquids and quantum phase transitions,''  arXiv:1110.3814 [hep-th].  %%CITATION = ARXIV:1110.3814;%%

%%%% Three-charge-BH
%\cite{Cvetic:1999xp}
\bibitem{Cvetic:1999xp}
  M.~Cvetic {\it et al.},
  ``Embedding AdS black holes in ten and eleven dimensions,''
  Nucl.\ Phys.\  B {\bf 558}, 96 (1999)
  [arXiv:hep-th/9903214].
  %%CITATION = NUPHA,B558,96;%%


%%%% Fermi liquid analysis
%\cite{Wu:2011cy}
\bibitem{Wu:2011cy}
  J.~P.~Wu,
  ``Some properties of the holographic fermions in an extremal charged
  dilatonic black hole,''
  Phys.\ Rev.\  D {\bf 84}, 064008 (2011)
  [arXiv:1108.6134 [hep-th]].
  %%CITATION = PHRVA,D84,064008;%%

%\cite{Edalati:2010ww}
\bibitem{Edalati:2010ww}
  M.~Edalati, R.~G.~Leigh and P.~W.~Phillips,
  ``Dynamically Generated Mott Gap from Holography,''
  Phys.\ Rev.\ Lett.\  {\bf 106}, 091602 (2011)
  [arXiv:1010.3238 [hep-th]].
  %%CITATION = PRLTA,106,091602;%%

%%%% Dipole coupling in RN BH

%\cite{Edalati:2010ge}
\bibitem{Edalati:2010ge}
  M.~Edalati, R.~G.~Leigh, K.~W.~Lo and P.~W.~Phillips,
  ``Dynamical Gap and Cuprate-like Physics from Holography,''
  Phys.\ Rev.\  D {\bf 83}, 046012 (2011)
  [arXiv:1012.3751 [hep-th]].
  %%CITATION = PHRVA,D83,046012;%%

%%%% Recent papers

%\cite{Guarrera:2011my}
\bibitem{Guarrera:2011my}
  D.~Guarrera and J.~McGreevy,
  ``Holographic Fermi surfaces and bulk dipole couplings,''
  arXiv:1102.3908 [hep-th].
  %%CITATION = ARXIV:1102.3908;%%

%\cite{Li:2011nz}
\bibitem{Li:2011nz}
  W.~-J.~Li and H.~Zhang,
  ``Holographic non-relativistic fermionic fixed point and bulk dipole coupling,''
  JHEP {\bf 1111}, 018 (2011)
  [arXiv:1110.4559 [hep-th]].
  %%CITATION = ARXIV:1110.4559;%%

%\cite{Wu:2012fk}
\bibitem{Wu:2012fk}
  J.~P.~Wu and H.~B.~Zeng,
  ``Dynamic gap from holographic fermions in charged dilaton black branes,''
  arXiv:1201.2485 [hep-th].
  %%CITATION = ARXIV:1201.2485;%%

%\cite{Li:2011sh}
\bibitem{Li:2011sh}
  W.~-J.~Li, R.~Meyer and H.~-b.~Zhang,
  ``Holographic non-relativistic fermionic fixed point by the charged dilatonic black hole,'' arXiv:1111.3783 [hep-th].  %%CITATION = ARXIV:1111.3783;%%

%\cite{Iizuka:2011hg}
\bibitem{Iizuka:2011hg}
  N.~Iizuka, N.~Kundu, P.~Narayan and S.~P.~Trivedi,
  ``Holographic Fermi and Non-Fermi Liquids with Transitions in Dilaton
  Gravity,''
  JHEP {\bf 1201}, 094 (2012)
  [arXiv:1105.1162 [hep-th]].
  %%CITATION = JHEPA,1201,094;%%

\bibitem{future work}
Work in progress
%\cite{Gursoy:2011gz}
\bibitem{Gursoy:2011gz}
  U.~Gursoy, E.~Plauschinn, H.~Stoof and S.~Vandoren,
  ``Holography and ARPES Sum-Rules,''
  arXiv:1112.5074 [hep-th].
  %%CITATION = ARXIV:1112.5074;%%

%\cite{Alishahiha:2012nm}
\bibitem{Alishahiha:2012nm}
  M.~Alishahiha, M.~R.~Mohammadi Mozaffar and A.~Mollabashi,
  ``Fermions on Lifshitz Background,''
  arXiv:1201.1764 [hep-th].
  %%CITATION = ARXIV:1201.1764;%%

%\cite{Fang:2012pw}
\bibitem{Fang:2012pw}
  L.~Q.~Fang, X.~H.~Ge and X.~M.~Kuang,
  ``Holographic fermions in charged Lifshitz theory,''
  arXiv:1201.3832 [hep-th].
  %%CITATION = ARXIV:1201.3832;%%
\end{thebibliography}
\end{document}